\documentclass[twocolumn,trackchanges]{aastex701}
\usepackage{natbib}  
\usepackage{threeparttable}
\usepackage{amsmath}
\usepackage{textcomp} 
\usepackage{siunitx}
\usepackage{booktabs}
\usepackage{adjustbox}
\usepackage{array}
\usepackage{siunitx}

\begin{document}

\title{The atmospheric extinction curve at Lenghu site}

\author[orcid=0009-0004-7361-3323,gname=Jin-Sheng,sname=Qiu]{Jin-Sheng Qiu\textsuperscript{\dag}}
\affiliation{Department of Astronomy, University of Science and Technology of China, Hefei 230026, China}
\affiliation{School of Astronomy and Space Sciences, University of Science and Technology of China, Hefei 230026, China}
\email{qjs003@mail.ustc.edu.cn}

\begingroup
\renewcommand{\thefootnote}{\dag} 
\footnotetext{Co-first authors} 
\endgroup


\author[orcid=0009-0009-3592-0214]{Xiao-Hui Xu\textsuperscript{\dag}}
\affiliation{Department of Astronomy, University of Science and Technology of China, Hefei 230026, China}
\affiliation{School of Astronomy and Space Sciences, University of Science and Technology of China, Hefei 230026, China}
\email{xiaohuixu@mail.ustc.edu.cn}

\correspondingauthor{Qing-Feng Zhu: zhuqf@ustc.edu.cn}
\author[orcid=0000-0003-0694-8946]{Qing-Feng Zhu}
\affiliation{Department of Astronomy, University of Science and Technology of China, Hefei 230026, China}
\affiliation{School of Astronomy and Space Sciences, University of Science and Technology of China, Hefei 230026, China}
\email{zhuqf@ustc.edu.cn}

\author{Xu Kong}
\affiliation{Department of Astronomy, University of Science and Technology of China, Hefei 230026, China}
\affiliation{School of Astronomy and Space Sciences, University of Science and Technology of China, Hefei 230026, China}
\email{}

\author{Ting-Gui Wang}
\affiliation{Department of Astronomy, University of Science and Technology of China, Hefei 230026, China}
\affiliation{School of Astronomy and Space Sciences, University of Science and Technology of China, Hefei 230026, China}
\email{}

\author{Lu-Lu Fan}
\affiliation{Department of Astronomy, University of Science and Technology of China, Hefei 230026, China}
\affiliation{School of Astronomy and Space Sciences, University of Science and Technology of China, Hefei 230026, China}
\affiliation{College of Physics, Guizhou University, Guiyang 550025, China}
\email{}

\author{Yong-Quan Xue}
\affiliation{Department of Astronomy, University of Science and Technology of China, Hefei 230026, China}
\affiliation{School of Astronomy and Space Sciences, University of Science and Technology of China, Hefei 230026, China}
\email{}

\author{Ji-An Jiang}
\affiliation{Department of Astronomy, University of Science and Technology of China, Hefei 230026, China}
\affiliation{National Astronomical Observatory of Japan, 2-21-1 Osawa, Mitaka, Tokyo 181-8588, Japan}
\email{}

\author{Zheng Lou}
\affiliation{Purple Mountain Observatory, Chinese Academy of Sciences, Nanjing 210023, China}
\email{}

\author{Xu Zhou}
\affiliation{National Astronomical Observatories, Chinese Academy of Sciences, Beijing 100101, China}
\email{}

\author{Xu-Zhi Li}
\affiliation{School of Mathematics and Physics, Anqing Normal University, Anqing 246133, China}
\affiliation{Institute of Astronomy and Astrophysics, Anqing Normal University, Anqing 246133, China}
\email{}

\author{Bo-Jun Tao}
\affiliation{Department of Astronomy, University of Science and Technology of China, Hefei 230026, China}
\affiliation{School of Astronomy and Space Sciences, University of Science and Technology of China, Hefei 230026, China}
\email{}

\author{Jun-Han Zhao}
\affiliation{Department of Astronomy, University of Science and Technology of China, Hefei 230026, China}
\affiliation{School of Astronomy and Space Sciences, University of Science and Technology of China, Hefei 230026, China}
\email{}

\author{Zhi-Yong Pu}
\affiliation{Department of Astronomy, University of Science and Technology of China, Hefei 230026, China}
\affiliation{School of Astronomy and Space Sciences, University of Science and Technology of China, Hefei 230026, China}
\email{}

\begin{abstract}

In this study, we use a dedicated spectroscopic telescope to carry out low-resolution measurements of the optical atmospheric extinction curve at  Lenghu astronomical site in Qinghai Province, China. Observations of A0-type stars are conducted over multiple nights between 2024 and 2026, covering airmasses from 1.0 to 2.0 and wavelengths in the range of 400 to 800 nm. We derive the extinction curve for the Lenghu site and compare it with those from Mauna Kea and Cerro Paranal. 
The results confirm that Lenghu site offers outstanding atmospheric conditions for astronomy,  making it highly suitable for optical astronomical observations.

\end{abstract}




\section{Introduction} 

Lenghu observatory at Saishiteng Mountain of Qinghai Province is a fast developing astronomical site in China due to its favorable atmospheric conditions, according to the long-term site monitoring conducted by \cite{2021Deng}. Statistical analyses show that the fraction of photometric nights per year in Lenghu exceeds 70\% \citep{2022Gao}. The seeing has a median value of 0.80 arcseconds and is better than 0.63 arcseconds during 25\% of the night time \citep{2021Deng,2023Zhu}. The level of light pollution is low and the measured average night-sky background brightness is approximately 22.0 mag arcsec$^{-2}$ in V-band on the moonless night \citep{2021Deng}. Due to the promising observational conditions at the Lenghu site, an increasing number of telescopes are being deployed or are already in operation. Among them, the 2.5-meter Wide Field Survey Telescope (WFST)  is a large field of view (6.5-square-degree field), photometric (g-band limited magnitude of 24 mag with 90s exposure), and time-domain survey telescope \citep{WFST, 2023Lei,2016Lou}.

Even at a particularly excellent site, the influence of Earth’s atmosphere on ground-based astronomical observations is non-negligible. Lenghu is no exception. As starlight traverses the atmosphere, it is attenuated by the scattering and absorption. This combined process is called extinction. The atmospheric extinction is primarily governed by three physical mechanisms \citep{1975Hayes,1977Houghton}. Rayleigh scattering arises from atmospheric atoms and molecules whose sizes are equivalent to or smaller than optical wavelengths $\lambda$.  The extinction due to Rayleigh scattering scales roughly as $\lambda^{-4}$, making it the main component of extinction at short wavelengths. 
Aerosol extinction originates from suspended solid and liquid particles (such as soot particles, sea-salt particles) in the atmosphere and generally follows a power-law dependence on wavelength \citep{1929A,1964A}. The power-law exponent is related to the column density and particle size distribution of the aerosol mixture \citep{1992Reimann, 2013Buton}. Additionally, molecules in the atmosphere produce absorption features at specific wavelengths or bands. For example, ozone exhibits the Chappuis absorption band between 500 nm and 700 nm \citep{1880chappuis}. Oxygen, water vapor and other molecules exhibit narrow telluric absorption features, mainly above 650nm \citep{2011Patat}. Oxygen causes two prominent absorption bands centered at approximately 687 nm and 760 nm, commonly referred to as the $O_2B$ and $O_2A$ bands \citep{2003Hinkle}, respectively. Water vapor exhibits a distinct absorption band near 720 nm \citep{2003Hinkle}. 

For an astronomical  site, the accurate measurement of the atmospheric extinction is essential for precision photometry and flux calibration \citep{2011Patat}.
Previous atmospheric extinction monitoring programs have been established at world-class observatory sites such as Cerro Paranal in Chile and Mauna Kea in Hawaii. 
\cite{1987Krisciunas} measured the atmospheric extinction at Mauna Kea in the B band ($\sim$440 nm) and V band ($\sim$550 nm) using the University of Hawaii 2.2 m telescope (UH88) and the 3.6 m Canada–France–Hawaii Telescope (CFHT). 
\cite{2013Buton}  conducted high-precision extinction measurements over a broad optical wavelength range ($320-970$ nm) using the SuperNova Integral Field Spectrograph (SNIFS), mounted on the UH88 telescope. 
\cite{2011Patat} used the FOcal Reducer/low-dispersion Spectrograph (FORS1), mounted at the ESO Kueyen 8.2 m telescope, to precisely measure the atmospheric extinction properties at Cerro Paranal. The above studies greatly benefit observational researches carried out at these sites and serve as a useful reference for our work.

The aim of this study is to measure the atmospheric extinction for Lenghu site. We deploy a dedicated spectroscopic telescope to carry out  extinction measurements. In section \ref{sec2}, we introduce the instrument for our study and the observations of the target stars. In section \ref{sec3}, we present the data reduction and the calibration methods. In section \ref{sec4}, we show our results, which are compared with the theoretical MODTRAN models and the extinction curves of Mauna Kea and Cerro Paranal.  Section \ref{sec5} shows the conclusions of our studies.

\section{Observations} 
\label{sec2}
\subsection{Instrument} 

In this study, we use a small aperture telescope and a grism spectrograph to measure the atmospheric extinction curve. The system includes a GSO 200mm Ritchey–Chrétien (RC) reflecting telescope with a focal length of 1600mm \citep{1927Ritchey}. Attached to the telescope’s focal plane is an ALPY-600  grism spectrograph \citep{ALPY}. The grism has a line density of 600 mm$^{-1}$, which disperses optical light with a resolving power of R $\sim$ 600.
The slit is 23 $\mu m$ (3") wide and 3 mm long. Spectral data are recorded using an Atik460EX CCD camera with a sensor size of 2749 × 2199 pixels. A secondary Atik314L+ CCD camera provides the real-time guiding through the slit viewing optices.  The basic parameters of the spectroscopic telescope are listed in Table \ref{spectrograph telescope}. The entire setup is mounted on a robust equatorial mount and operated remotely.  We use control softwares N.I.N.A\footnote{Stefan Berg et al, https://nighttime-imaging.eu/} to point the telescope to the selected star. Then we use TheSkyX\footnote{https://www.bisque.com/downloads/thesky-user-guide/} to identify the star though the image analysis. Finally we use PHD Guiding\footnote{Craig Stark et al, https://openphdguiding.org/} to center the star on the slit and guide. MaxIm DL Pro6\footnote{Doug George et al, https://www.diffractionlimited.com} is used to record the spectral data from CCD camera.

\begin{table}[htbp]
\centering
\caption{The basic parameters of the spectroscopic telescope}
\label{spectrograph telescope}
\begin{threeparttable}
\begin{tabular}{lc}
\hline
Parameters & Values \\
\hline
Aperture & 200 mm  \\
Focal length & 1600 mm \\
Focal plate scale & 128.9"/mm \\
Slit width & 23$\mu$m (3")\\
Slit length & 3mm \\
Spectral resolution & 600 @650nm\\

\hline
\end{tabular}

\end{threeparttable}
\end{table}

\subsection{Observational Targets}

\newcounter{tempcnt}
\setcounter{tempcnt}{5}
\begin{table*}[htbp]
\centering
\caption{The A0 observational stars sample}
\label{A0V observational stars}
\begin{threeparttable}
\begin{tabular}{ccccccc}
\hline
Stellar name &  Photovisual magnitude & Exposure(s) & Dist(pc) & RA(deg) & DEC(deg) & Stellar ID\\
\hline
HD 172167  & 0.14 & 2 & 7.68 & 279.240 & 38.769 & 1\\

HD 358  & 2.15 & 10 & 29.74 & 2.089 & 29.090 & 2\\

HD 186882  & 2.97 & 10 & 47.48 & 296.255 & 45.126 & 3\\

HD 40183  & 2.07 & 10 & 24.87 & 89.886 & 44.943 & 4\\

HD 112185  & 1.68 & 10 & 25.31 & 193.494 & 55.958 & 5\\

HD 95418  & 2.44 & 10 & 25.91 & 165.454 & 56.380 & 6\\

HD 47105  & 1.93 & 10 & 33.51 & 99.418  & 16.399  & 7\\

HD 123299  & 3.64 & 20 & 79.87 & 211.106 & 64.372 & 8\\

HD 177724 & 3.02 & 20 & 26.16 & 286.349 & 13.868 & 9\\

HD 139006  & 2.31 & 10 & 23.67 & 233.680 & 26.716 & 10\\

HD 103287 & 2.54 & 10 & 25.50 & 178.460 & 53.694 &11\\

HD 161868 & 3.74 & 20 & 29.75 & 266.980 & 2.714 & 12\\

HD 198001 & 3.83 & 20 & 74.84 & 311.928 & -9.499 & 13\\

HD 191692 & 3.37 & 20 & 70.08 & 302.814 & -0.820 & 14\\

HD 218045 & 2.57 & 10 & 40.88 & 346.194 & 15.206 & 15\\

\hline
\end{tabular}
\end{threeparttable}
\end{table*}

The observations are carried out mostly during photometric, dark nights to ensure the atmospheric  stability and minimal sky background contamination. The spectral type A0 stars are used for our study, as \cite{2011Patat}. The spectral energy distributions of A0 stars exhibit  relatively smooth continuum across optical  wavelengths \citep{1988Wade}. The dominant line features are hydrogen Balmer lines, which are narrow, deep and located at precisely known wavelengths. This allows the stellar absorption lines to be reliably masked. High-precision empirical templates of A0 spectra are publicly available, such as from the Pickles library \citep{1998Pickles} or the HST CALSPEC database\footnote{https://www.stsci.edu/hst/instrumentation/reference-data-for-calibration-and-tools/astronomical-catalogs/calspec}\citep[]{2014BohlinPASP,2014BohlinAJ}.
Our A0 target stars are selected from the Henry Draper catalog \citep{1993Cannon}, and the observational stars are listed in Table \ref{A0V observational stars}.
It includes the following information: the photovisual magnitude, exposure time, the heliocentric distance, RA, DEC, and the identification number assigned by us. We select the exposure time that ensures the high-quality continuum spectra are obtained but none pixel is saturated. Usually, 5 exposures for the same star are taken to improve the final signal-to-noise ratio through averaging.

A0 target stars of different altitude angles, or airmasses are selected for observations in each night. As described in \cite{1989Kasten}, the airmass (X) can be calculated as Eq \ref{eq1}:

\begin{equation}
    X(h) = \frac{1}{\cos [90 - h(^\circ)] + 0.5057 \times [6.0799 + h(^\circ)]^{-1.6364}}
    \label{eq1}
\end{equation}
where $h$ is the altitude angle of a star. On each observing night, we observe stars of various altitude angles so that the airmass is distributed between $X=1$ and $X=2$. Observations are conducted such that the selected stars are roughly uniformly distributed along the airmass. The observation dates, the range of airmass and the observed stars are shown in Table \ref{Observation days}. Our observations are carried out in two modes:
(1) observing multiple stars at different airmasses, and
(2) continuously tracking a single star as its airmass changes with time (2024-05-11, 2024-05-13, 2025-09-28). It turns out two modes produce equivarlent results.

\begin{table}[htbp]
\centering
\caption{The observation dates and the airmass ranges}
\label{Observation days}
\begin{threeparttable}
\begin{tabular}{lcc}
\hline
Date & Airmass Range & Stellar ID \\
\hline
2024-05-11 & 1.42 - 2.09 & 1 \\
2024-05-13 & 1.00 - 1.47 & 1 \\
2024-11-29 & 1.08 - 1.69 & 1, 2, 3\\
2024-12-02 & 1.04 - 2.02 & 1, 2, 3\\
2024-12-04 & 1.12 - 2.17 & 1, 4\\
2025-04-28 & 1.05 - 1.93 & 5, 6, 7, 8\\
2025-04-30 & 1.05 - 1.93 & 5, 6, 7, 8\\
2025-06-15 & 1.03 - 1.81 & 1, 3, 5, 9, 10, 11\\
2025-09-22 & 1.07 - 1.90 & 1, 10\\
2025-09-28 & 1.02 - 1.96 & 1\\
2025-10-20 & 1.05 - 1.88 & 1, 3, 9, 12, 13, 14\\
2025-10-26 & 1.07 - 1.85 & 1, 3, 9, 13, 15\\
2025-11-24 & 1.02 - 1.92 & 1, 2, 3, 9, 15\\
2025-12-15 & 1.01 - 1.73 & 2, 3, 15\\
2025-12-16 & 1.02 - 2.06 & 1, 2, 3, 15\\
2025-12-23 & 1.02 - 1.91 & 2, 3, 15\\
2026-01-11 & 1.08 - 2.05 & 2, 4, 15 \\
\hline
\end{tabular}

\end{threeparttable}
\end{table}

\section{Data Reduction and Calibration} \label{sec3}

The raw spectral data undergo typical reduction and calibration processes. The reduction process begins with bias subtraction and dark current correction, which remove fixed-pattern bias and thermal electronic signal from the detector. The bias and dark frames are taken at the beginning and the end of each night. Bias frames are taken with zero-second exposures, while dark frames are taken with exposure times matched to those of the stellar observations. Multiple bias and dark fames are taken each time to reduce the statistical fluctuation.

\subsection{Spectral Calibration}

The calibration process includes wavelength
calibration and flux calibration. We calibrate the wavelength using emission lines from a mercury (Hg) lamp.
We fit the spectral lines using a single Gaussian profile for isolated lines and a double Gaussian profile for double lines to obtain the central pixel positions. 
We match the central pixel positions to the reference wavelengths of Hg lines from the NIST Atomic Spectra Database \footnote{\url{https://www.nist.gov/pml/atomic-spectra-database}} \citep{NIST_ASD}. 
The pixel-wavelength relation is derived by fitting a third-order polynomial.


To obtain spectral flux distribution, we correct for the instrumental response, which includes telescope mirror reflectivity, spectrograph grating efficiency, and CCD quantum efficiency \citep{2010Burke}.
Since our measurements are limited in visible wavelength range, we perform an empirical calibration using a xenon (Xe) lamp light source. 
The relative spectral intensity distribution of the xenon lamp is measured in laboratories at the University of Science and Technology of China (USTC) and the Anhui Institute of Optics and Fine Mechanics, Chinese Academy of Sciences (AIOFM). 
We also use a portable commercial fiber spectrometer to measure the spectrum of the lamp before the observations to make sure the measured spectra are consistent with the lab results.
The spectrograph of the same xenon lamp spectrograph is observed using  our telescope-spectrograph through a diffuse light box at the Lenghu site. The observed spectral image serves as the flat-field for the spectrograph observations. 
By taking the ratio of the observed flat-field spectrum to the laboratory Xenon lamp spectrum, we derive the system response function $R(\lambda)$ to calibrate the observed spectral flux. The responses  from different CCD rows are  highly consistent over the wavelength range $400-800$ nm. The Xe lamp signal is low beyond this range and the derived response curves are less reliable. Therefore, only spectral data in the $400-800$ nm range are considered in subsequent analysis.

\begin{figure*}
    \hspace{-50pt}
    \begin{adjustbox}{width=1.2\linewidth}
        \includegraphics{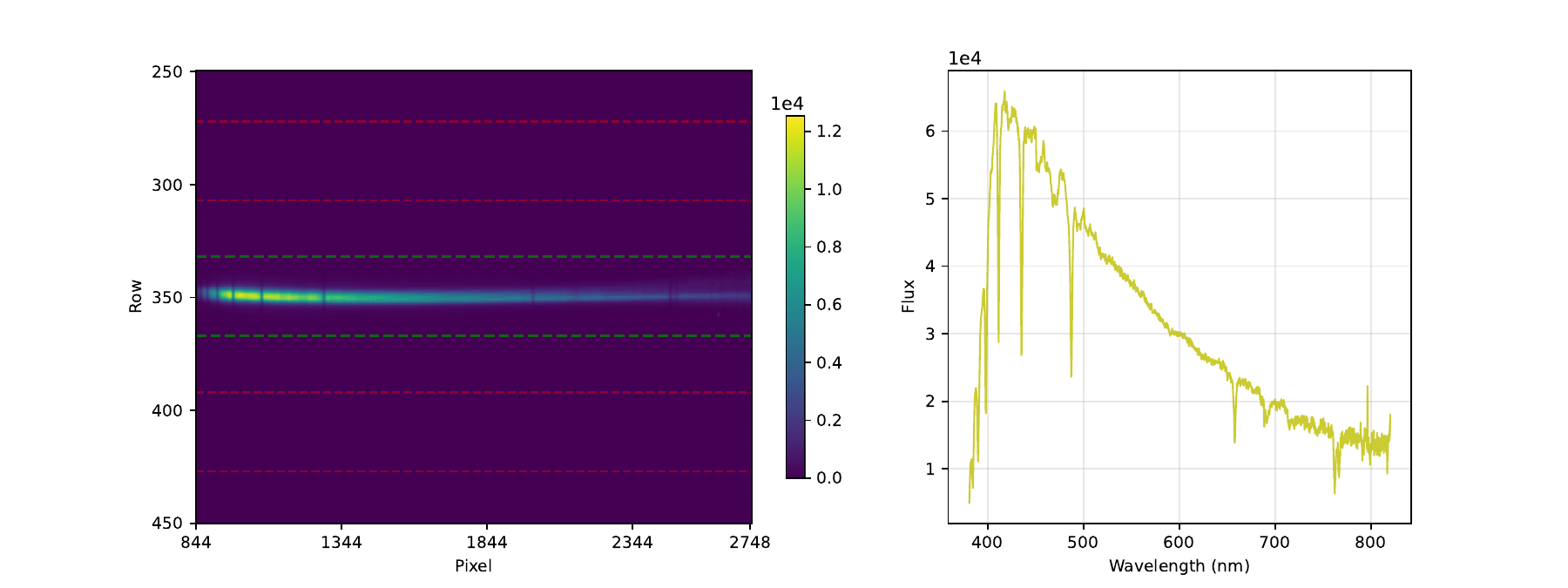}
    \end{adjustbox}
    \caption{
    The left panel shows the two-dimensional the stellar spectrum image after calibration. The target stellar spectrum is extracted from the region between the two green dashed lines (35 pixels). The sky background is measured by averaging the flux in the regions above and below the stellar spectrum (35 pixels, between the red dashed lines).  The right panel displays the one-dimensional extracted spectrum, obtained by summing the flux column within the extraction region.}
    \label{fig2}
\end{figure*}

Finally, the sky background is subtracted following the normal slit spectrograph background subtraction procedure. Typically, 35 pixels are used to extract stellar spectrum. This is illustrated in the left panel of Figure \ref{fig2}, where the spectrum is extracted from the region between the two green dashed lines in the calibrated stellar spectrum. The regions between the red dashed lines (also 35 pixels) are used for background spectrum extraction. 

Following the standard spectroscopic data reduction steps, including bias and dark current subtraction, pixel-wavelength calibration, instrumental response correction and sky background subtraction, the raw spectra image are processed and one-dimensional calibrated spectra  are extracted by summing pixel values perpendicular to the dispersion direction. One example of the extracted spectrum is shown in right panel of Figure \ref{fig2}. 

\begin{figure*}
    \centering
    \includegraphics[width=1\linewidth]{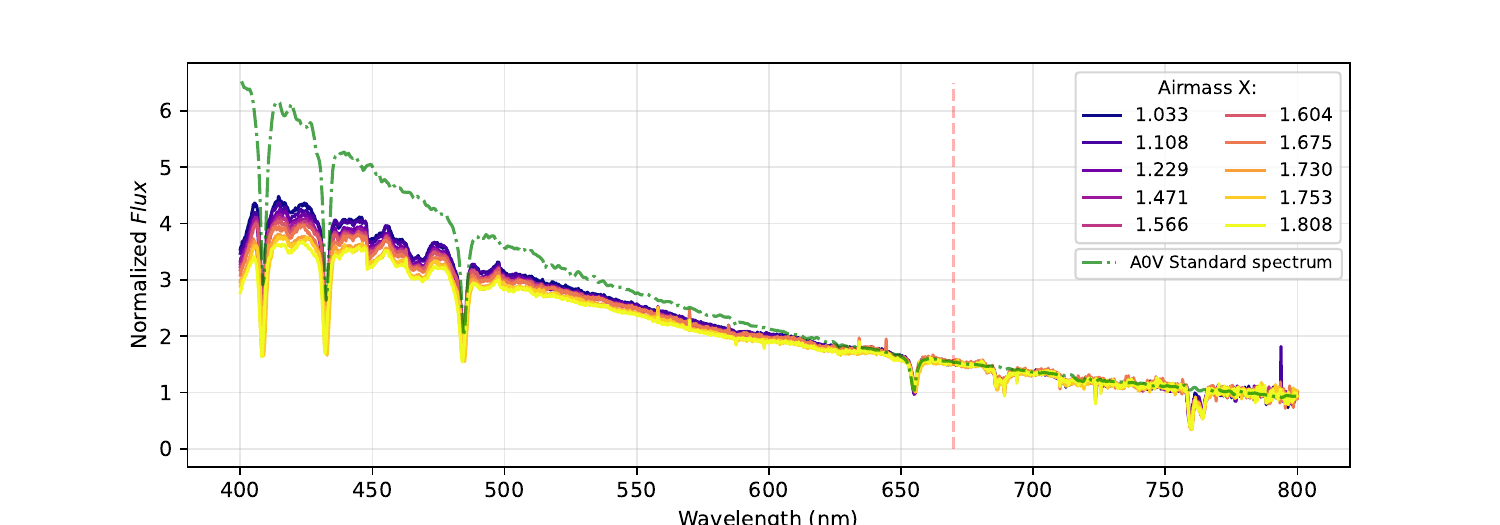}
    \caption{The image shows the normalized observed spectra at different airmasses, with different colors representing different airmass values. The green dash-dotted line presents the normalized A0 standard star spectrum for comparison. The red dashed vertical line indicates the wavelength of 670 nm. Beyond this wavelength, multiple absorption features are evident in the observed spectra that are absent in the A0 standard spectrum.
}
    \label{fig3}
\end{figure*}
Figure \ref{fig3} shows the example of the normalized spectra of target stars observed at different airmasses in a photometric night (after calibration). Each spectrum is normalized using the flux within the wavelength interval $775-785$ nm, a region relatively free of telluric absorption features. The color gradient represents the airmass variation. The bluer color corresponds to the smaller airmass. The green dash-dotted line shows the standard A0-type stellar spectrum from the Pickles library \citep{1998Pickles}. As shown in Figure \ref{fig3}, for wavelengths greater than 670 nm, the observed spectrum exhibits absorption-lines features that are not present in the standard A0 stellar spectrum. These features are caused by the absorption of atmospheric oxygen and water molecules, and their positions are consistent with those identified in previous studies \citep{2003Hinkle}. We retain the atmospheric absorption lines and mask the A0 stellar absorption lines in our analysis.


\subsection{Atmospheric Dispersion Correction}

Since our telescope does not have a field rotater, the slit is not aligned with the meridian direction of the target during our observations. As noted in previous studies \citep{1982Filippenko,2011Patat}, atmospheric dispersion would introduce the bias in the measurements of atmospheric extinction coefficients. Due to the wavelength-dependent refraction in the atmosphere, stellar images are stretched along meridians: light of shorter wavelength is refracted toward the zenith, while light of longer wavelength is refracted toward opposite direction. When the slit is misaligned with the meridian, the fraction of spectral flux entering the slit varies with wavelength, causing the bias of the derived extinction coefficients.

We construct a simple model to correct for the flux loss caused by slit misalignment. We divide the wavelength range from 400 nm to 800 nm into 10 nm bins and simulate the offset of the stellar image at each wavelength bin relative to a reference wavelength of 500 nm. The reference wavelength is chosen based on the sensitivity of the guiding camera which has a peak QE (quantum efficiency) at around 500 nm. The offsets $\Delta R(\lambda)$ are computed according to the atmospheric dispersion formula \citep{1982Filippenko}:

\begin{equation}
    \Delta R(\lambda)(") = 206265 \times [n_{P, T}(\lambda) - n_{P, T}(500)] \times \tan z
    \label{eq5}
\end{equation}
where $n_{P, T}(\lambda)$ is the refractive index of the atmosphere at wavelength $\lambda$. $P$ and $T$ are the pressure and temperature of observational site. For $P$ and $T$, we adopt the typical values at the Lenghu site: 460 mmHg and \SI{-5}{\celsius}. $z$ is the zenith angle of observation target. 

\begin{figure*}
\centering
\includegraphics[width=1\textwidth]{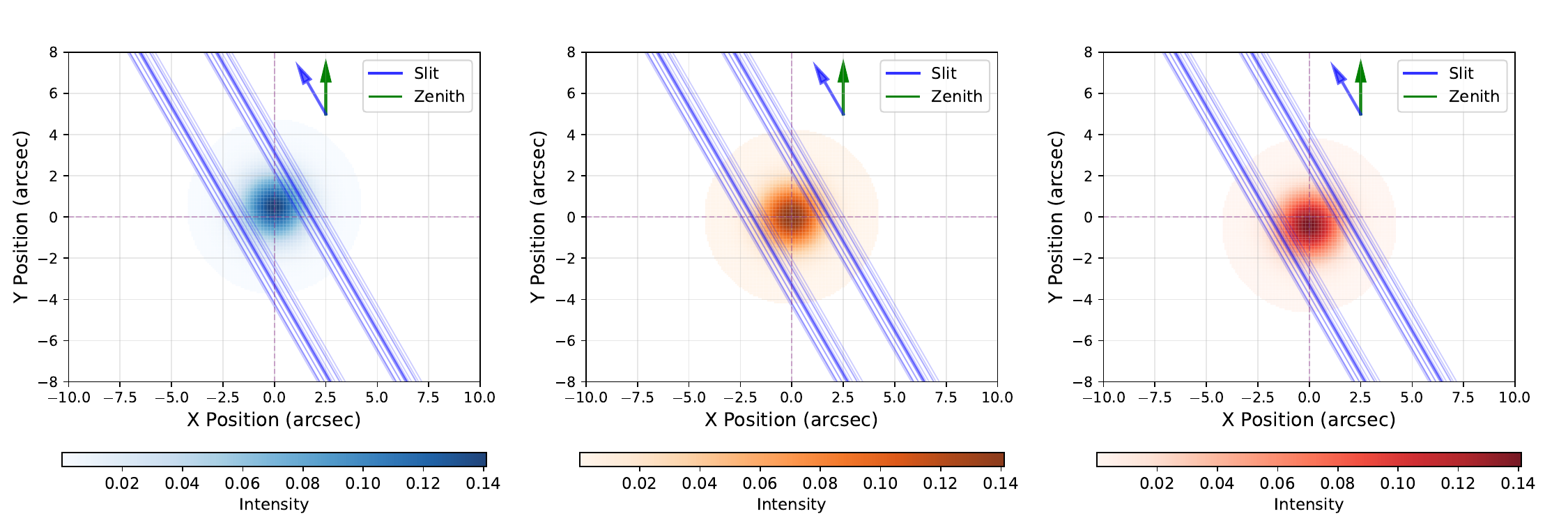}
\caption{The stellar images at three different wavelengths (blue for 400 nm, orange for 500 nm, red for 700 nm), which are Gaussian PSFs with a FWHM of 2.5". The positive direction of the y-axis represents the zenith direction, as indicated by the green arrow. The slit blue stripes are fixed at 3" width and the slit centers are randomly sampled from a Gaussian distribution ($\sigma=0.8"$) around the center position.}
\label{Appendix-fig1}
\end{figure*}

As an illustrative example for our atmospheric dispersion model, we show stellar images at three wavelengths—400 nm, 500 nm (the reference wavelength, centered at (0, 0)), and 700 nm—in Figure \ref{Appendix-fig1}. We assume that stellar image follows a Gaussian profile with a FWHM (full-width at half maximum) of 2.5". This is the typical stellar image size by analyzing the slit viewing images. The offsets $\Delta R(\lambda)$ along y-axis is calculated using Eq \ref{eq5}, with a zenith angle $z = 45^\circ$ adopted as an illustrative example. 

The positive direction of the y-axis represents the zenith direction, as indicated by the green arrow in the figure. As an illustrative example, we set the angle between the slit and the zenith direction to 30$^\circ$.
Due to telescope guiding errors and atmospheric seeing, the actual center of the stellar image in the slit varies randomly. We simulate this by sampling the slit center from a Gaussian distribution perpendicular to the slit direction with a standard deviation $\sigma$ of 0.8". We show 20 random simulations of the slit position with blue stripes in the figure and calculate the average fraction of stellar flux that passes through the slit at each wavelength bin.



For each observation at a given airmass, we calculate the angle between the slit and the zenith direction. Using our atmospheric dispersion model, we derive the wavelength-dependent slit throughput and apply it as a correction for the slit misaligned effect. 

\subsection{Atmospheric Extinction Derivation}

Given the proximity of our selected A0 standards (the distances are typically a few tens of parsecs and E(B-V) from 3D dustmap is zero \citep{2019Green}), interstellar extinction is negligible. The calibrated spectrum is related to the intrinsic stellar spectrum  through atmospheric extinction, as the function \citep{1988Wade,1992Reimann, 2013Buton}:

\begin{equation}
    F_{cali}(\lambda; X) = F_{int}(\lambda) \times 10^{-0.4k(\lambda)X}
    \label{eq3}
\end{equation}
where k($\lambda$) is the atmospheric extinction coefficient, and X is the airmass of the observation.
We compute the ratio between the calibrated stellar spectrum (after telescope responce calibration and the slit misaligned correction) and the standard spectrum of A0 star. The ratio 
\begin{equation}
    T(\lambda; X)=\frac{F_{cali}(\lambda; X)}{F_{int}(\lambda)}, 
\end{equation}
directly relates to the atmospheric transmission. We compute the average transmission in individual wavelength bin for different airmasses. We adopt a non-uniform wavelength bins across the $400-800$ nm range as follow:

\begin{equation}
\Delta \lambda =
\begin{cases}
10~\mathrm{nm}, \lambda < 650~\mathrm{nm}, \\
5~\mathrm{nm}, \lambda > 650~\mathrm{nm} 
\end{cases}
\label{eq:binning}
\end{equation}

\begin{figure}
    \centering
    \hspace*{-0.85cm}
    \includegraphics[width=1.1\linewidth]{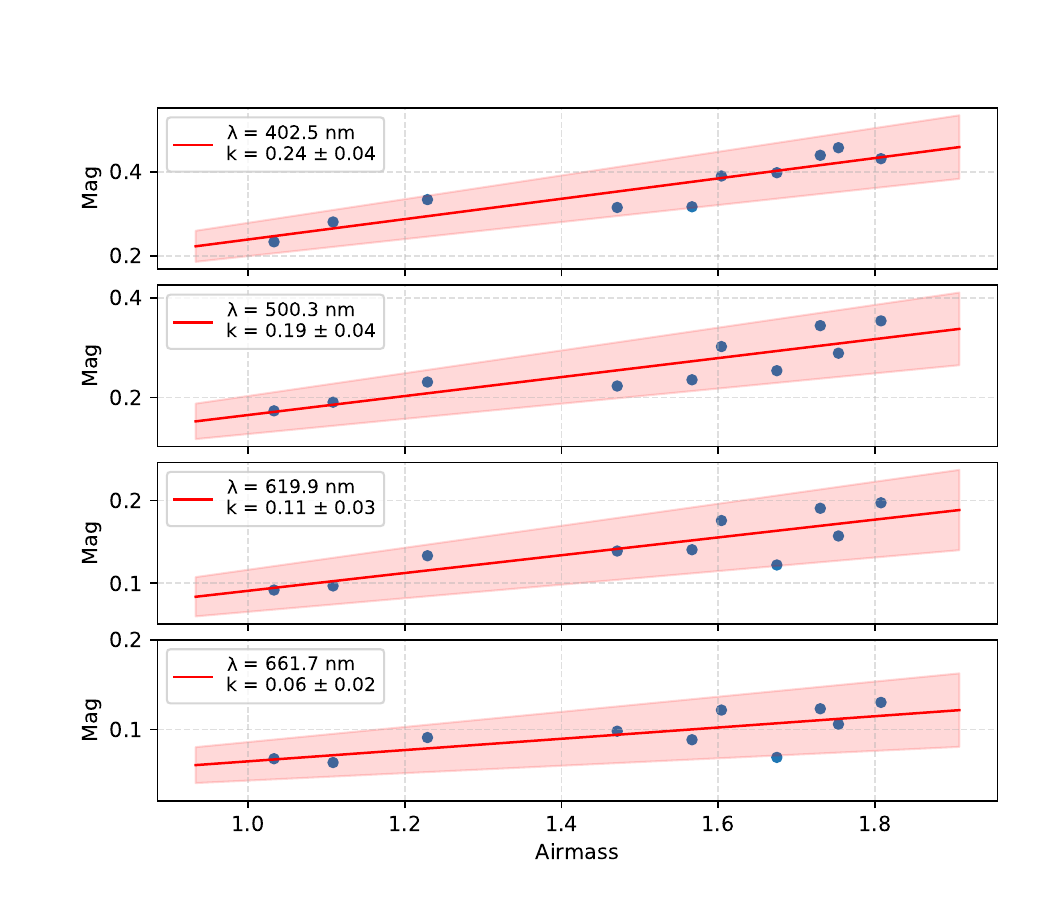}
    \caption{The figure displays the relationship between extinction magnitudes and airmasses at four selected wavelengths. Blue dots represent the observed data, the red line denotes the best linear fit, and the light red shaded region indicates the $\pm$ 1 standard deviation interval. The corresponding wavelengths and atmospheric extinctions k are labeled at upper-left corner. }
    \label{fig4}
\end{figure}

Within each wavelength bin i, we integrate the ratio and obtain the extinction value \citep{2011Patat}:
\begin{equation}
    \begin{split}
    T_i(X) = \frac{\int_{\lambda_i}^{\lambda_{i}+\Delta \lambda} T(\lambda; X) d\lambda}{\Delta \lambda} \\
    A_i(X) = -2.5 \log_{10} [T_i(X)]
    \end{split}
\end{equation}

For each wavelength bin, we plot $A_{i}(X)$ against X as shown in the Figure \ref{fig4}. According to \cite{2011Patat, 2013Buton}, the atmospheric extinction follows a linear relation: 
\begin{equation}
    A_i(X) = b_i + k_i X
\end{equation}

We perform a linear least-squares fit to the (X, $A_i$) data points, yielding $k_{i}$ and its uncertainty. $k_i$ is the extinction coefficient at the wavelength. The normalization constant and the slit flux loss effects are contained in the fitting constant $b_i$ and are less relevant for our analysis. Repeating this procedure for all wavelength bins produces the extinction coefficient curve k($\lambda$) over $400 - 800$ nm.

\section{Results}
\label{sec4}

\begin{figure*}
\centering
\includegraphics[width=1\textwidth]{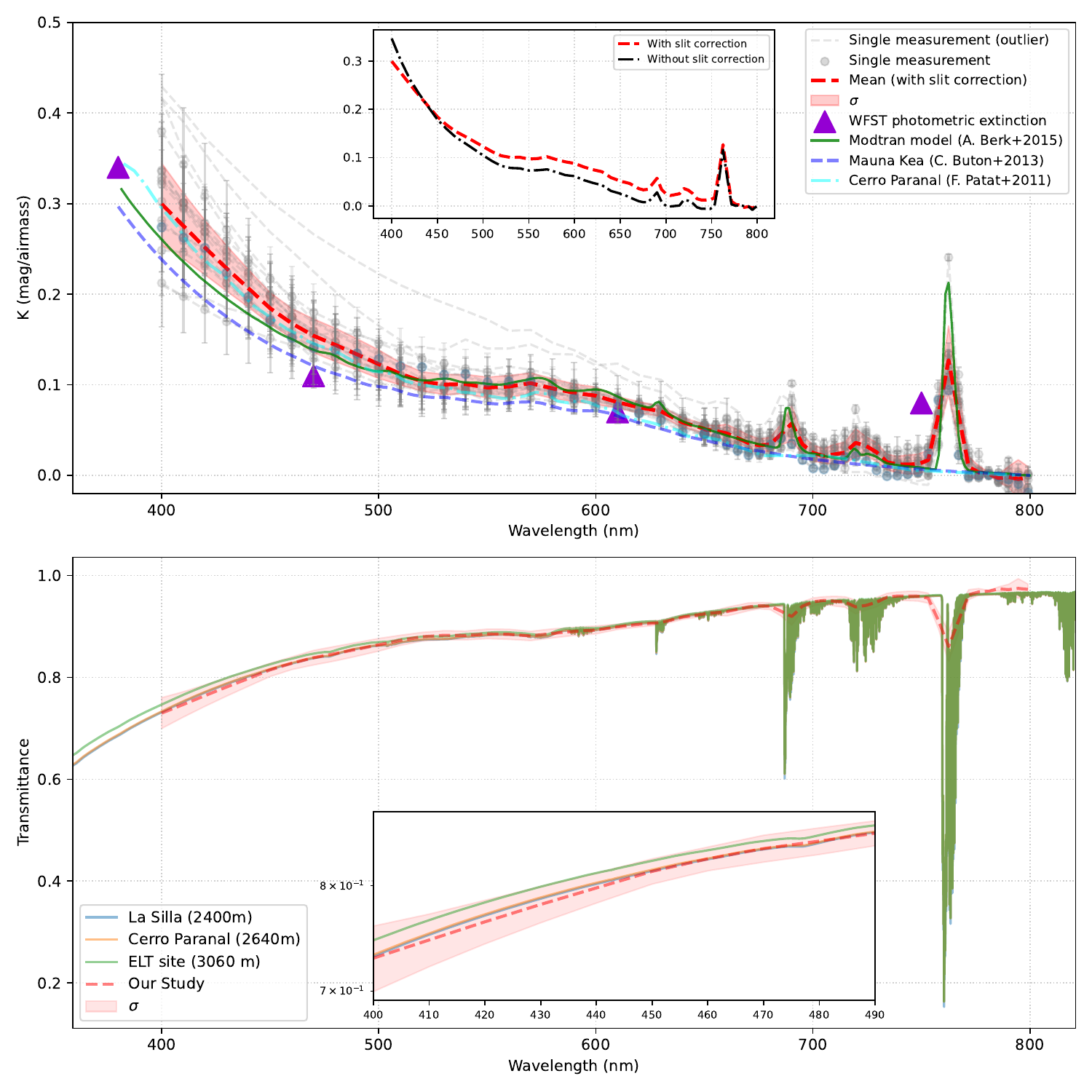}
\caption{In upper panel, the gray dotted-lines with errorbars represent individual nights' measurements, and gray dashed-lines indicate data excluded from averaging due to significant deviations. The red dashed line shows the average extinction curve, with a light-red shaded region for the standard deviation. For comparison, the green solid line shows the extinction curve of best-fitting MODTRAN model, the blue dashed line represents the extinction curve of Mauna Kea, the cyan dash-dotted line corresponds to the extinction curve of Cerro Paranal, and the purple triangles indicate photometric extinctions from WFST \citep[in prep]{Cai2026}. The inset plot highlights the difference between the extinction coefficients with (red dashed) and without (black dash-dotted) slit effect correction.
Bottom panel shows the atmospheric transmittance curves at the zenith for the Lenghu site and other observatory sites from SkyCalc. The inset plot is a local magnification.}
\label{fig5}
\end{figure*}

The upper panel of Figure \ref{fig5} shows the atmospheric extinction curves from our study, along with results from other researches. The colored dotted lines represent results from different nights. The red dashed line indicates the average atmospheric extinction. Gray dashed lines denote data excluded from the averaging process due to significant deviations identified via visual inspection.
For comparison, the inset plot in the upper panel of Figure \ref{fig5} shows the mean atmospheric extinction curves with and without slit effect correction, represented by a red dashed line and a black dash-dotted line, respectively.
The average atmospheric extinction data are presented in Table \ref{Atmospheric Extinction Result}.

Atmospheric extinction coefficients in the u, g, r and i bands have been calculated with photometric data taken by the WFST at Lenghu site \citep[in prep]{Cai2026}.
For comparison, we plot the photometric results from WFST as purple triangles in the upper panel of Figure \ref{fig5}.

\begin{table}[htbp]
    \centering
    \caption{Mean atmospheric extinction result}
    \label{Atmospheric Extinction Result}
    \begin{threeparttable}
    \begin{tabular}{rrrrrr}
        \hline
        $\lambda$ & $k$ & $\sigma_{k}$ & $\lambda$ & $k$ & $\sigma_{k}$ \\
        \hline
        400 & 0.300 & 0.045 & 655 & 0.049 & 0.006 \\
        410 & 0.275 & 0.036 & 660 & 0.046 & 0.010 \\
        420 & 0.252 & 0.030 & 665 & 0.041 & 0.009 \\
        430 & 0.229 & 0.025 & 670 & 0.036 & 0.009 \\
        440 & 0.207 & 0.021 & 675 & 0.033 & 0.007 \\
        450 & 0.184 & 0.017 & 680 & 0.034 & 0.008 \\
        460 & 0.168 & 0.016 & 685 & 0.045 & 0.011 \\
        470 & 0.154 & 0.019 & 690 & 0.057 & 0.020 \\
        480 & 0.144 & 0.018 & 695 & 0.034 & 0.010 \\
        490 & 0.134 & 0.017 & 700 & 0.025 & 0.009 \\
        500 & 0.123 & 0.016 & 705 & 0.022 & 0.008 \\
        510 & 0.112 & 0.012 & 710 & 0.023 & 0.011 \\
        520 & 0.104 & 0.011 & 715 & 0.025 & 0.012 \\
        530 & 0.100 & 0.011 & 720 & 0.036 & 0.016 \\
        540 & 0.101 & 0.011 & 725 & 0.033 & 0.010 \\
        550 & 0.097 & 0.012 & 730 & 0.026 & 0.011 \\
        560 & 0.098 & 0.012 & 735 & 0.015 & 0.009 \\
        570 & 0.102 & 0.011 & 740 & 0.012 & 0.009 \\
        580 & 0.096 & 0.011 & 745 & 0.012 & 0.010 \\
        590 & 0.091 & 0.010 & 750 & 0.012 & 0.011 \\
        600 & 0.088 & 0.009 & 755 & 0.017 & 0.011 \\
        610 & 0.081 & 0.006 & 760 & 0.066 & 0.019 \\
        620 & 0.075 & 0.006 & 765 & 0.127 & 0.038 \\
        630 & 0.071 & 0.006 & 770 & 0.066 & 0.030 \\
        640 & 0.059 & 0.005 & 775 & 0.009 & 0.011 \\
        650 & 0.051 & 0.005 & 780 & 0.003 & 0.007 \\
        &       &       & 790 & 0.002 & 0.012 \\
        \hline
    \end{tabular}
    \begin{tablenotes}
    \small
    \item[Note]
       $\lambda$ in nm; $k$ and $\sigma_{k}$ in mag/airmass.
    \end{tablenotes}
    \end{threeparttable}
\end{table}

\subsection{MODTRAN Model}
MODTRAN\footnote{http://modtran.spectral.com}(MODerate resolution atmospheric TRANsmission)  is a widely used radiative transfer model developed and maintained by the Spectral Sciences, Inc.(SSI) and Air Force Research Laboratory (AFRL)\citep{2015Berk}. 
 MODTRAN accounts for the key atmospheric radiative transfer processes through the Earth's atmosphere, including molecular absorption, aerosol scattering, and Rayleigh scattering. The range of wavelength covers from the ultraviolet to the far-infrared (approximately $0.2 - 200$ $\mu$m) with a spectral resolution on the order of \AA.

 From the MODTRAN website, we download atmospheric transmittance curves for different zenith angles and derive the atmospheric extinction coefficient curve $k_{model}(\lambda)$. We select the models with the parameters matching the Gobi environment conditions of Lenghu site. In Figure \ref{fig5} upper panel, the green solid line represents the best-fit Modtran model corresponding to our observational results. The associated parameters are: Ozone Column = 0.36 atm-cm, Aerosol Model='DESERT'.

 \subsection{Comparisons}

\cite{2011Patat} measured the atmospheric extinction curve at Cerro Paranal in Chile and investigated the temporal variability.
\cite{2013Buton} conducted a seven-year program of atmospheric extinction monitoring at Mauna Kea, accumulating a large dataset and deriving a robust extinction  curve. We compare our extinction  curve from the Saishiteng Mountain at Lenghu site with the results from Cerro Paranal and Mauna Kea, as shown in Figure \ref{fig5} upper panel. The cyan dash-dotted line represents the Cerro Paranal extinction curve from \cite{2011Patat}. The blue dashed line represents the Mauna Kea extinction curve from \cite{2013Buton}. The results show that the atmospheric extinction at the Lenghu site is in agreement with that of Cerro Paranal. As shown in Figure \ref{fig5}, the extinction coefficient of the wavelength of 400 nm at Mauna Kea is approximately 0.05 mag/airmass lower than that in Lenghu. It reflects the differences in atmospheric conditions between Mauna Kea and Lenghu sites.

Specifically, we compare the atmospheric transmittance curves at the zenith for the Lenghu site and other observatory sites in the bottom panel of Figure \ref{fig5}. The atmospheric transmittance curves at other sites are obtained from SkyCalc\footnote{\url{https://www.eso.org/observing/etc/bin/gen/form?INS.MODE=swspectr+INS.NAME=SKYCALC}} (Version 2.0.9). SkyCalc is developed by ESO (European Southern Observatory) based on the Cerro Paranal Advanced Sky Model \citep{2012Noll,2013Jones,2014Moehler}. SkyCalc  provides the atmospheric transmittance curves at three astronomical sites: Cerro Paranal (2400 m), La Silla (2600 m) and the Extremely Large Telescope (ELT) site (3060 m) as shown in the bottom panel of Figure \ref{fig5}. The result shows that the atmospheric transmittance curve of the Lenghu site is similar to those of the three sites. This is because all three sites are located in the Atacama Desert, whose conditions closely resemble those of the Gobi Desert at Lenghu site \citep{2023Lei}.

\section{Conclusions}
\label{sec5}

In this work, we present the first  atmospheric extinction curve ($400-800$ nm) measurements for  Lenghu site, Qinghai, based on systematic spectroscopic observations of A0 stars.  
The curve clearly shows the characteristics of Rayleigh scattering, the ozone Chappuis bands ($500-700$ nm), $O_2A$ and $O_2B$ absorption bands (760 nm and 687 nm), and $H_2O$ absorption band near 720 nm. There is strong consistency between our results and the preliminary photometric extinction results based on the WFST at the Lenghu observatory site.
Furthermore, we compare the atmospheric extinction at Lenghu with those measured at the world-class observatories at Mauna Kea, Cerro Paranal and La Silla, etc. The results indicate that Lenghu exhibits extinction levels closer to those high altitude desert sites, such as Cerro Paranal and La Silla. This may be due to the fundamental differences in the atmospheric composition between the desert and the sea areas. These results again suggest that the Lenghu Saishiteng mountain site is an outstanding astronomical site in terms of the atmospheric extinction. 

Our study provides essential atmospheric characterization for the Lenghu astronomical site. The work helps the accurate flux calibration for the scientific observations of current and future facilities at Lenghu site.


\begin{acknowledgments}
This work is supported by the National Key Research and Development Program of China (2023YFA1608100). We acknowledge the support of the Strategic Priority Research Program of the Chinese Academy of Sciences (grant NO. XDB0550300). We sincerely thank Prof.Keqiang Qiu, Prof.Wenxin Huang, Dr.Rucheng Dai, and Dr.Zhuo Wang for their invaluable assistance in the calibration of the Xe lamp light source. We sincerely thank Mr.Weiguo Zhang and Mr.Ruoyu Zhou for their invaluable assistance in the construction of the spectroscopic telescope and their help during the observations. We thank Dr.Zhen Wan and Mr.Minxuan Cai for the photometric extinction data from WFST. We also thank Dr.Zelin Xu for their assistance during the observations.
\end{acknowledgments}

\software{astropy \citep{2013A&A...558A..33A,2018AJ....156..123A,2022ApJ...935..167A}, MaxIm DL Pro 6, N.I.N.A, PHD Guiding, TheSkyX}

\bibliography{my}{}
\bibliographystyle{aasjournalv7}



\end{document}